\documentclass[aps,prx,twocolumn,superscriptaddress]{revtex4-2}
\usepackage{graphicx}
\usepackage{xcolor}
\usepackage[normalem]{ulem} 
\usepackage{array}

\begin{document}


\title{Monolithic Polarizing Circular Dielectric Gratings on Bulk Substrates for Improved Photon Collection from InAs Quantum Dots}


\author{Ryan A. DeCrescent}
\email{ryan.decrescent@nist.gov}
\affiliation{National Institute of Standards and Technology, Boulder, Colorado 80305, USA}
\author{Zixuan Wang}
\affiliation{National Institute of Standards and Technology, Boulder, Colorado 80305, USA}
\affiliation{Department of Physics, University of Colorado, Boulder, CO 80309, USA}
\author{Poolad Imany}
\affiliation{National Institute of Standards and Technology, Boulder, Colorado 80305, USA}
\affiliation{Department of Physics, University of Colorado, Boulder, CO 80309, USA}
\author{Sae Woo Nam}
\affiliation{National Institute of Standards and Technology, Boulder, Colorado 80305, USA}
\author{Richard P. Mirin}
\affiliation{National Institute of Standards and Technology, Boulder, Colorado 80305, USA}
\author{Kevin L. Silverman}
\email{kevin.silverman@nist.gov}
\affiliation{National Institute of Standards and Technology, Boulder, Colorado 80305, USA}

\date{\today}

\begin{abstract}
III-V semiconductor quantum dots (QDs) are near-ideal and versatile single-photon sources. Because of the capacity for monolithic integration with photonic structures as well as optoelectronic and optomechanical systems, they are proving useful in an increasingly broad application space. Here, we develop monolithic circular dielectric gratings on bulk substrates -- as opposed to suspended or wafer-bonded substrates -- for greatly improved photon collection from InAs quantum dots. The structures utilize a unique two-tiered distributed Bragg reflector (DBR) structure for vertical electric field confinement over a broad angular range. Opposing ``openings" in the cavities induce strongly polarized QD luminescence without harming collection efficiencies. We describe how measured enhancements depend critically on the choice of collection optics. This is important to consider when evaluating the performance of any photonic structure that concentrates farfield emission intensity. Our cavity designs are useful for integrating QDs with other quantum systems that require bulk substrates, such as surface acoustic wave phonons.
\end{abstract}


\maketitle

\section{Introduction}
III-V semiconductor quantum dots (QDs) are recognized as quintessential solid-state single-photon sources for quantum photonic technologies \cite{senellart_high-performance_2017}. They emit on-demand indistinguishable single photons at gigahertz rates with nearly lifetime-limited spectral linewidths \cite{monniello_indistinguishable_2014, lobl_narrow_2017, tomm_bright_2021}. Their charge states can be deterministically controlled with simple semiconductor gate structures, and their resonance frequencies can be Stark tuned within the same device layout \cite{hogele_voltage-controlled_2004, lobl_narrow_2017}. These features --- combined with the possibility of monolithic integration --- offer tremendous opportunities for interfacing III-V QDs with other two-level systems and for incorporating them into larger hybrid systems and circuits such as optoelectronic or optomechanical systems \cite{tsuchimoto_large-bandwidth_2022, decrescent_large_2022, imany_quantum_2022, buhler_-chip_2022, papon_independent_2023, finazzer_-chip_2023, larocque_tunable_2023}. 

One universal obstacle for the implementation of QD light sources is due to the relatively large refractive index mismatch between the host medium (e.g., GaAs) and vacuum. The majority of the photons generated in a bulk material experience total internal reflection at the semiconductor-vacuum interface, ultimately limiting photon collection efficiencies to $\lesssim$1\% when using vertical collection and external optics. A wide variety of photonic structures have been developed to efficiently interface with QDs for both on-chip and free-space applications \cite{lodahl_interfacing_2015}. Some examples include photonic crystal waveguides \cite{hughes_extrinsic_2005, lund-hansen_experimental_2008}, ridge waveguides and ring resonators \cite{luxmoore_interfacing_2013, dusanowski_purcell-enhanced_2020}, and microdisk resonators \cite{michler_quantum_2000, peter_exciton-photon_2005}, all of which are particularly useful for on-chip routing of photons to and from QDs. For free-space applications, micropillar cavities \cite{reithmaier_strong_2004, somaschi_near-optimal_2016}, photonic crystal cavities \cite{strauf_self-tuned_2006}, circular grating resonators \cite{davanco_circular_2011, liu_solid-state_2019} and open tuneable microcavities based on distributed Bragg reflectors (DBRs) \cite{najer_gated_2019} have been demonstrated. Such architectures are often designed to optimize a specific metric, e.g., photon collection efficiency, strong exciton-photon coupling, optical coherence times, or total brightness. Though this is suitable for pure photonic applications, these structures often cannot be immediately incorporated into larger hybrid structures where the QD needs to interact well with another system. 

An example hybrid system that encounters this challenge is a microwave-to-optical transducer based on InAs QDs and surface acoustic wave (SAW) resonators \cite{imany_quantum_2022, decrescent_large_2022}. This technology requires optimized electrical, mechanical and optical structures to be co-located while minimally afflicting the other subsystems. Recent work has shown remarkable success, but poor optical collection from the QDs was a significant source of total end-to-end efficiency losses. Specifically, a bare GaAs surface is ideal for high-quality-factor SAW resonators, but leads to poor photon collection from the QD. On the other hand, most previously developed photonic structures for optimal photon collection will strongly scatter the SAW field, reducing mechanical quality factors, or change the mode shape completely. This hybrid system thus requests a photonic structure monolithically incorporated into a bulk substrate while minimally perturbing the SAW strain field. The current work is largely motivated by this goal, but our cavity designs may be useful for any application where quantum emitters must be embedded in bulk substrates, such as integrating with bulk acoustic resonators \cite{machado_generation_2019} or similar vertical acoustic microcavities \cite{kuznetsov_microcavity_2023}.

Our designs are based on circular dielectric gratings, or ``bullseye cavities", which have been shown to greatly improve vertical extraction of single photons emitted from QDs \cite{davanco_circular_2011, liu_solid-state_2019}. Previous work used suspended membranes or wafer-bonded III-V layers to vertically confine the optical fields so that the emitted photons readily interact with the radial bullseye structure \cite{davanco_circular_2011, liu_solid-state_2019}. Here, we show how to achieve effective vertical field confinement by using a two-tiered DBR structure. This offers several potential advantages over suspending or wafer-bonding approaches, including fabrication ease and maintaining larger distances between the QD and etched surfaces. In order to make these structures compatible with SAW resonators, we open the optical cavities on two sides so that focused SAWs can propagate through them with minimal scattering. This opening also creates an optical anisotropy that leads to highly polarized luminescence while negligibly affecting optical performance for the cavity-polarized emission mode. We measure $\approx$100$\times$ photon collection improvements from QDs in our bullseye cavities when compared to unstructured regions on the same substrates; calculations suggest that this corresponds to $\approx$30$\times$ improvements compared to a traditional DBR structure. Motivated by our findings, we quantitatively describe the critical role that external collection optics have on measured collection enhancements (Appendix B). That concept is applicable to any photonic structure that concentrates far-field emission intensity and should be considered carefully when evaluating device performance.

\section{Design and fabrication}
\begin{figure}
      \includegraphics[width=0.9\columnwidth]{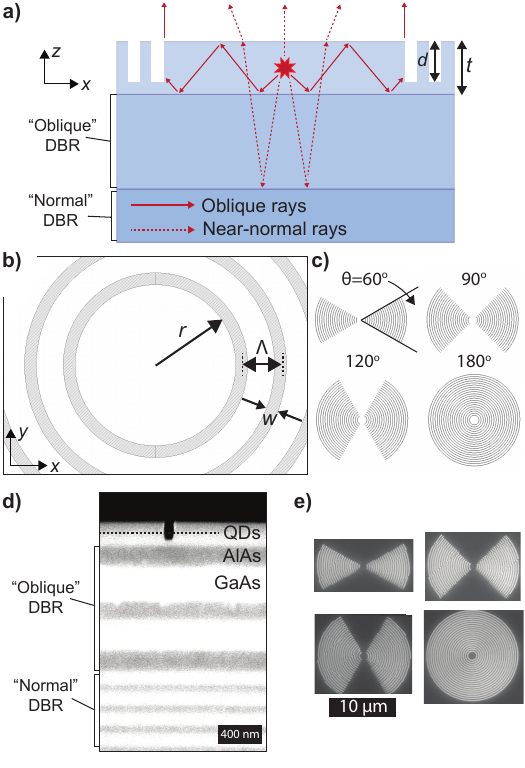}
	\caption{	
	(\textbf{a}) Cross-sectional schematic ($x$-$z$ plane) of the device. Etched grooves (white regions), ``normal DBR" (dark blue) and ``oblique DBR" (light blue) regions are designated. Red lines illustrate how light emitted from a QD (red star) at various angles interacts with the structure. 
	(\textbf{b}) In-plane ($x$-$y$) structure of the center of the device, illustrated to scale. Gray corresponds to etched regions. Several design parameters are designated in panels a and b. 
	(\textbf{c}) Full in-plane structures of four devices, illustrated to scale, differing only by a cavity enclosure angle $\theta$. 
	(\textbf{d}) Cross-sectional SEM of the wafer structure. White regions: GaAs. Gray regions: AlAs. Distinct normal DBR and oblique DBR regions are designated. A single etched groove is apparent. QDs are grown at the center of the top GaAs slab (black dotted line). 
	(\textbf{e}) Plan-view SEM images of four distinct devices, corresponding to the four devices in panel c. The 10 $\mu$m scale bar applies to panels e and c. 
	}
\label{fig:1}
\end{figure}

Our devices (Fig. \ref{fig:1}a) consist of a GaAs slab (thickness $t$) above two distinct DBR regions. The lower DBR consists of 22 periods of AlAs/GaAs and is designed to reflect normal-incidence light (``normal DBR"). The upper DBR consists of 2.5 periods of relatively thick AlAs/GaAs layers and is designed to reflect oblique-incidence light at angles around 63$^\circ$ (``oblique DBR"). InAs QDs are grown at the center of the upper GaAs slab. Circular grooves with depth d are etched into the resulting heterostructure, defining the optical bullseye cavity. The normal DBR reflects light that would otherwise be lost into the bulk substrate, but is only effective within an angular range spanning approximately 20$^\circ$ around normal incidence (Appendix A). The upper oblique DBR is intended to emulate a slab waveguide so that light emitted at larger angles (55$^\circ$ to 70$^\circ$) readily interacts with the circular grating. The basic radial geometry is defined by three parameters (Fig \ref{fig:1}b): the center radius ($r$), trench periodicity ($\Lambda$), and trench width ($w$). Design parameters are optimized by estimating device performance using commercial finite-difference time-domain software. Final design parameters are specified in Table I and in the text when relevant.

\begin{table*}
\centering
\begin{tabular}{| c | m{6em} | m{6em} | m{6em} | m{7em} | m{6em} | m{6em} | m{7em} |}
\hline
Parameter & Normal DBR layer thicknesses & Oblique DBR layer thicknesses & GaAs slab thickness, $t$ & Grating etch depth, $d$ & Grating periodicity, $\Lambda$ & Trench width, $w$ & Cavity center radius, $r$ \\
\hline
Value & 81.0 nm (AlAs) / 69.2 nm (GaAs) & 188.2 nm (AlAs) / 188.2 nm (GaAs) & 172 nm & 0.83$t$=142.7 nm & 325 nm & 0.21$\Lambda$=68 nm & 2.025$\Lambda$=58 nm \\
\hline
\end{tabular}
\caption{Optimized geometrical parameters.}
\end{table*}

Finally, we symmetrically open the cavity trenches such that opposing etched minor arcs span an angle $\theta$$<$180$^\circ$. Three partially enclosed cavities with cavity enclosure angles $\theta$=60$^\circ$, 90$^\circ$, and 120$^\circ$ are illustrated in Fig. \ref{fig:1}c. In this geometry, $y$-oriented electric dipoles are expected to interact with the grating while $x$-oriented dipoles are expected to be only weakly affected. Fully enclosed cavities with $\theta$=180$^\circ$ are expected to show polarization-independent performance.

Samples are grown via molecular beam epitaxy and then deposited with a sputtered SiO$_2$ hard mask. Circular grating trenches are defined by electron-beam lithography and subsequently etched via reactive-ion etching. The hard mask is then removed by hydrofluoric acid. Fig. \ref{fig:1}d shows a cross-sectional scanning electron micrograph (SEM) of a fabricated calibration structure; distinct oblique DBR and normal DBR regions, as well as a single etched groove, are easily identified. Fig. \ref{fig:1}e shows plan-view SEMs of four cavities with $\theta$=60$^\circ$, 90$^\circ$, 120$^\circ$, and 180$^\circ$ (corresponding to structures illustrated in Fig. \ref{fig:1}c) during an intermediate fabrication step.  

\begin{figure}
      \includegraphics[width=0.9\columnwidth]{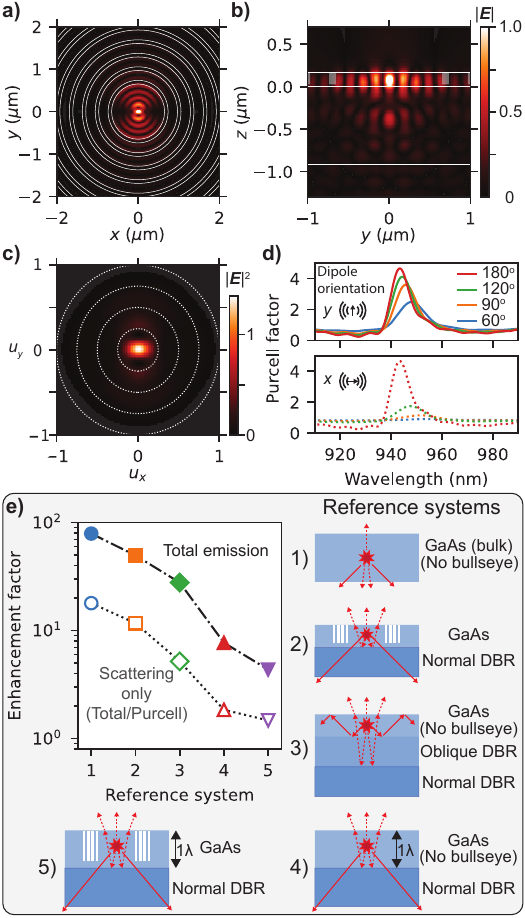}
	\caption{	
	(\textbf{a,b}) Calculated field magnitude ($|E|$) in the (a) $x$-$y$ plane and (b) $y$-$z$ plane when driving a fully enclosed cavity with an $x$-oriented electric dipole on resonance at 945 nm. In a, edges of the etched regions are designated by white circles. In b, boundaries between the DBRs and GaAs slab regions are designated by white horizontal lines and etched regions by gray rectangles. 
	(\textbf{c}) Farfield intensity $|E|^2$ calculated in vacuum (above the device) under the same conditions as in panels a and b. Dotted white circles indicate increments of NA=0.25.
	(\textbf{d}) Purcell spectra calculated for $y$-oriented (top panel) and $x$-oriented (bottom panel; dashed curves) dipoles. Different colors correspond to different cavity enclosure angles $\theta$ according to the legend. The system geometry (illustrated in each panel) is the same as in Fig. \ref{fig:1}c. 
	(\textbf{e}) Photon collection enhancements, relative to five different reference systems (illustrated at right and below), calculated for bullseye cavities. ``Total rate" includes both Purcell and geometrical enhancements. ``Scattering only" excludes changes due to the Purcell factor. (More complete explanations of these terms are provided in the main text.) In all cases, collected photons correspond to power contained in the farfield within an NA of 0.5.
	}
\label{fig:2}
\end{figure}

The optical bullseye cavities are designed to exhibit an $(n,l)$=(5,0) drumhead-like electromagnetic resonance ($n$ and $l$ are the radial and azimuthal quantum numbers of a circular resonator) at 945 nm. Numerically calculated electric field magnitude ($|E|$) profiles of this cavity resonance, excited by an $x$-oriented electric dipole, are illustrated in Figs. \ref{fig:2}a,b for two different plane cuts. These calculations show that the circular grating and double-DBR structures generate substantial in-plane (Fig. \ref{fig:2}a) and out-of-plane (Fig. \ref{fig:2}b) field confinement. Farfield calculations (Fig. \ref{fig:2}c) show that a majority of the optical power emitted into the vacuum above the device is contained within an angular range corresponding to a numerical aperture (NA) of 0.25. This directed emission is favorable when long-working-distance collection optics must be used, a scenario commonly encountered with optical cryostats. The cavities also theoretically provide modest Purcell emission rate enhancements of approximately 4 to 5 (Fig. \ref{fig:2}d). Here, we use the Purcell spectrum primarily to identify and quantify the cavity resonance. We also quantify the polarizing properties of partially enclosed cavities by comparing the Purcell spectra for $y$- and $x$-oriented dipoles. These spectra indicate cavity resonances with a typical bandwidth of 10 nm. 

For $y$-oriented dipoles (Fig. \ref{fig:2}d; top panel), the Purcell enhancement for the $\theta$=90$^\circ$ partially enclosed cavity is reduced by only $\approx$25\% with respect to the fully enclosed cavity ($\theta$=180$^\circ$). In contrast, for $x$-oriented dipoles, the Purcell enhancement nearly vanishes for the $\theta$=90$^\circ$ cavity. This is intuitive when considering the radiation patterns for the respective dipole orientations; the 90$^\circ$ enclosed cavity scatters a majority of the $y$-oriented dipole’s radiation field, but very little of the $x$-oriented dipole’s field. In fact, this remains true even for off-center dipoles, and polarization-dependent photon collection enhancements are thus expected to be somewhat robust against QD positioning and to exist regardless of the Purcell effect. Fig. \ref{fig:2}e quantifies these effects by comparing calculated photon collection rates from our optimized bullseye cavities to five reference systems (illustrated on the right side and bottom of panel e). The reference systems are as follows: 1) a bare (bulk) GaAs substrate with no bullseye grating; 2) our optimized bullseye geometry \emph{without} the upper oblique DBR; 3) our optimized bullseye cavity geometry \emph{without} the bullseye trenches; 4) a conventional DBR structure comprising a 1-$\lambda$ thick GaAs on a normal DBR; 5) the same as `Reference system 4' with additional bullseye trenches.

In all cases, collected photons correspond to farfield power contained within an NA of 0.5. Total ``rate enhancements" (Fig. \ref{fig:2}e; solid markers) are derived by directly comparing the calculated farfield power between the optimized bullseye cavity and reference systems. We expect roughly 100$\times$ and 10$\times$ total rate enhancements when compared to bare GaAs (blue; `Reference system 1') and a conventional DBR structure (red; `Reference system 4'). Photon “collection enhancements” (Fig. \ref{fig:2}e; open markers) --- arising purely from the redistribution of emitted photons due to coherent scattering --- are derived by normalizing the farfield power by the total radiated dipole power. For example, a single emitted photon from our device is approximately 20$\times$ more likely to be collected when again compared to a bare GaAs surface (blue), and approximately 2$\times$ more likely to be collected when compared to a conventional DBR structure (red). Importantly, our optimized structures are still expected to provide 5$\times$ (2$\times$) higher photon rates (collection) than a conventional DBR structure even when adding a bullseye grating to that structure (purple; `Reference system 5'). These results indicate that the collection improvement in our system largely originates from the circular gratings and to a lesser extent from the two-tiered DBR structure, although there is a synergy between these two components. In Appendix B, we describe how \emph{experimentally} observed collection enhancements depend on the collection NA and other details of the experimental apparatus.

\section{Experimental device characterization}
For initial experimental characterization, we fabricate the aforementioned devices on wafers grown with a relatively high QD density (approximately 10 QDs per $\mu$m$^2$). We perform photoluminescence (PL) measurements using a home-built fiber-coupled confocal microscope around an optical cryostat with the sample held at a temperature of approximately 5 K. QDs are optically excited by an 827 nm (nonresonant) pump laser focused to a nearly diffraction-limited spot at the sample surface through an objective with a nominal NA of 0.7. (The general importance of the collection optics is discussed in Appendix B.) PL is collected by the same objective, then coupled into a single-mode optical fiber. Reflected pump light is rejected with spectral filters. Polarization-dependent PL spectra are recorded by transmitting the collected PL through a linear polarizer before being coupled into fiber and counted on a CCD spectrometer. 

\begin{figure*}
      \includegraphics[width=0.9\linewidth]{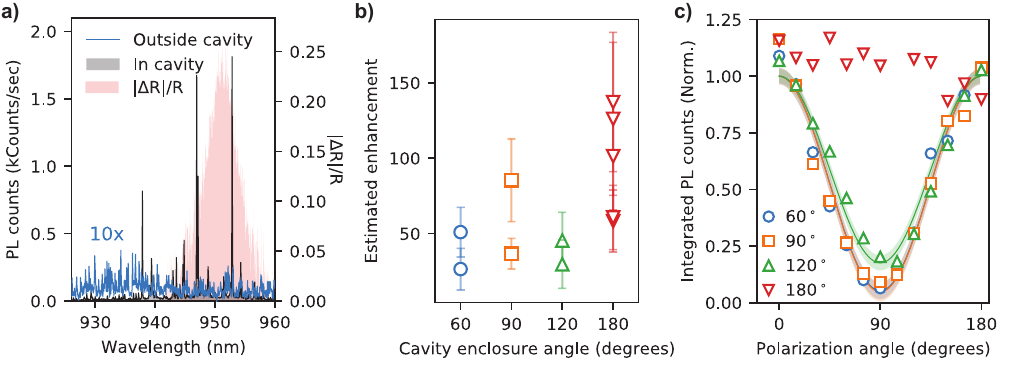}
	\caption{	
	(\textbf{a}) PL spectra recorded from QDs inside a $\theta$=90$^\circ$ cavity (black) and from a bare region immediately outside the same cavity (blue; multiplied by 10). The differential reflection spectrum $|\Delta R|/R$ (light-red filled region; right axis) from the same $\theta$=90$^\circ$ cavity is also shown. 
	(\textbf{b}) Estimated collection enhancement for a variety of cavities with different cavity enclosure angles $\theta$ (horizontal axis). All spectra in panels a and b were recorded with $y$ polarization. Enhancements are calculated by comparing the peak PL counts from a QD inside each cavity to characteristic peak PL counts from QDs immediately outside the cavity. Error bars represent uncertainties arising from the random brightness of each QD in the ensemble; specifically, they were derived by taking 10 distinct estimates for QD count rates outside the cavity within a 10 nm spectral range of the cavity's peak PL.
	(\textbf{c}) Experimental PL counts (open markers) of single QD emission lines as a function of collection polarization angle $\phi$ for cavities with different enclosure angles $\theta$ (specified in the legend). (One device of each $\theta$ was measured and plotted.) Fits to a sinusoidal angular variation are shown by solid curves. Data for $\theta$=60$^\circ$, 90$^\circ$, and 120$^\circ$ cavities are normalized to the fit value at polarization angle $\phi$=0$^\circ$; data for the $\theta$=180$^\circ$ cavity is normalized independently. 
	}
\label{fig:3}
\end{figure*}

Typical PL spectra are shown in Fig. \ref{fig:3}a, recorded from inside a partially enclosed cavity ($
\theta$=90$^\circ$, black filled spectrum) and from an unetched region immediately outside the cavity (blue, multiplied by 10). Spectra were recorded under identical pump and collection conditions, and correspond to $y$-polarized emission. At this QD density, approximately 10 to 15 QDs contribute to each recorded spectrum, yielding approximately 30 to 45 PL peaks spanning a wavelength range between 920 nm and 960 nm. (Each QD contributes 2 to 3 PL lines to each spectrum, originating from different charge states and exciton complexes.) Spectra recorded from inside the cavity show only a few ($\approx$3 to 6) intense PL lines with count rates approximately 20 to 50$\times$ higher than typical peak values recorded from bare regions. That is, the cavities allow spectral isolation and improved collection of just 2 to 3 QDs, likely those located near the cavity’s center. The enhanced peaks tend to lie within a 10 nm range around the cavity resonance, as verified by measuring each cavity’s reflection spectrum (e.g., Fig. \ref{fig:3}a; red filled region). This indicates that the improved collection indeed originates from a coherent scattering process associated with the designed cavity mode rather than random scattering from etched surfaces. The improved count rate agrees well with the calculated values illustrated in Fig. \ref{fig:2}e (green markers).

We perform a similar comparison on cavities with enclosure angles $\theta$=60$^\circ$, 90$^\circ$, 120$^\circ$, and 180$^\circ$. Fig. \ref{fig:3}b summarizes measured collection enhancements obtained from a variety of cavities with various values of $\Lambda$, $r$, and $w$ (not specified) and $\theta$ (horizontal axis). Due to the random nature of the brightness and position of each QD in the cavities, the estimated enhancement varies widely between cavities and shows no clear correlation with the enclosure angle $\theta$. (Error bars in Fig. 3b represent variations expected from the random brightness of QDs in the ensemble.) Nonetheless, typical PL enhancements are estimated to be around 30 to 50$\times$, with an upper estimate of approximately 140$\times$ from several of the fully enclosed cavities. We note that the magnitudes of these experimentally observed enhancements depend on the details of the experimental apparatus and are expected to be smaller for high-NA collection objectives (Appendix B).

A remarkable result is that PL enhancements for partially enclosed cavities are comparable to those of complete circular cavities --- differing only by a factor of $\approx$2 --- when collecting $y$-polarized PL. In contrast, $x$-polarized spectra from partially enclosed cavities resemble spectra recorded from unetched regions, indicating that $x$-polarized collected photons weakly interact with the etched structure. Polarization characteristics of the cavities are summarized in Fig. \ref{fig:3}c. In this analysis, photon counts from individual PL lines are evaluated as a function of polarization angle. For all partially enclosed cavities, the collected PL is a minimum for $x$-polarized collection (polarizer angle 90$^\circ$) and maximum for $y$-polarized collection (polarizer angles 0$^\circ$ and 180$^\circ$). As a result, the polarization contrast for both $\theta$=60$^\circ$ (blue circles) and $\theta$=90$^\circ$ (orange squares) cavities is approximately 20:1. The polarization contrast decreases to approximately 5:1 for the $\theta$=120$^\circ$ (green up triangles) cavity. Fully enclosed cavities (red down triangles) show no systematic polarization dependence; variations across angles likely arise from variations in our apparatus’ photon collection efficiencies as the polarizer is rotated.

\section{Discussion and conclusions}
We have detailed the design, fabrication, and optical characterization of circular dielectric gratings for improved photon collection from InAs QDs. As opposed to previous work which used suspended membranes or wafer-bonded III-V layers for effective vertical field confinement over a broad angular range \cite{davanco_circular_2011, liu_solid-state_2019}, we designed monolithic structures using a unique two-tiered DBR structure. We experimentally observe up to 140$\times$ photon collection enhancements when compared to unstructured regions on the same substrate. Based on numerical calculations, we thus anticipate approximately 30$\times$ better photon collection rates when compared to optimized conventional DBR structures. These collection enhancements are engineered to be strongly polarization dependent by simply truncating the grating structures angular extend around the QD. These anisotropic structures only weakly impact total collection enhancements.  

III-V QDs are very sensitive to localized charges within a several-hundred nanometer vicinity \cite{houel_probing_2012}. Consequently, for best performance, QDs should typically be kept away from interfaces where large surface defect densities are possible. One potential benefit of our monolithic bullseye gratings is that the nearest etched interfaces (namely, the bullseye trenches) are larger than 650 nm away from bullseye-centered QDs. In contrast, low-quality interfaces may be as close as 90 nm in suspended membranes \cite{davanco_circular_2011}, to 150 nm in wafer-bonded structures \cite{liu_solid-state_2019}. Our devices are also immediately compatible with conventional QD electrostatic p-i-n gating methods \cite{lobl_narrow_2017, najer_gated_2019}, and thus are well-suited for improving single-photon source efficiencies while retaining the desired low-noise characteristics of optimized III-V QDs. 

\section{Acknowledgements}
This research was performed while R.D. held an NRC Research Associateship award at NIST. 

\newpage
\section*{Appendix A: Design logic of two-tiered DBR}

\begin{figure}
      \includegraphics[width=0.8\columnwidth]{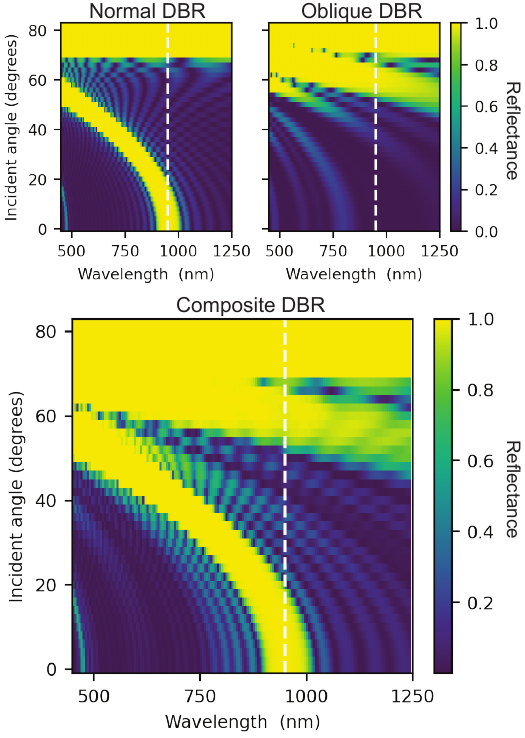}
      \caption{Calculated angle-dependent reflectance spectra from two distinct DBRs (top images) and the composite two-tiered DBR used in our fabricate devices (bottom image).}
\label{fig:A1}
\end{figure}

Figure \ref{fig:A1} shows calculated angle-dependent reflectance spectra from three distinct DBR structures. The top two images corresponding to the lower (``Normal DBR") and upper (``Oblique DBR") regions in our fabricated structures, calculated with 20 DBR periods in each structure. The bottom image corresponds to the fabricated compound two-tiered DBR system with only 2.5 periods in the oblique DBR. For these calculations, an $s$-polarized plane wave is incident from a semi-infinite GaAs layer. The vertical white dashed line indicates the design wavelength where high reflectance around both 0$^\circ$ and $\approx$63$^\circ$ are desired.

\section*{Appendix B: Compatibility between photonic structures and collection optics}
Our experimental setup uses a single-mode optical fiber to collect the QD’s emission. End-to-end collection efficiencies thus depend on first-lens collection efficiency and on proper fiber coupling. Generally, optics needed for best performance when measuring emission from a photonic structure differ from those needed without the photonic structure. For this reason, evaluating the performance of certain photonic structures depends on the larger optical setup and care should be taken when comparing results across systems. 

\begin{figure}
      \includegraphics[width=0.9\columnwidth]{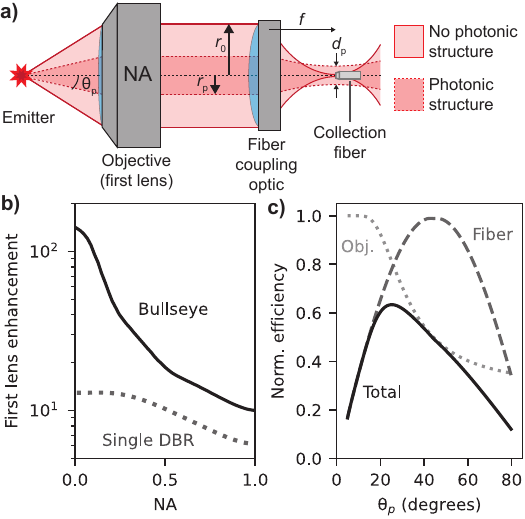}
	\caption{	
	(\textbf{a}) Schematic setup for collecting emitted photons into a single-mode optical fiber. The fiber-coupling optics were optimized for collection from an emitter without a photonic structure. With a photonic structure, emitted light is concentrated within a cone corresponding to divergence angle $\theta_p$. The objective collimates this collected light into a beam with radius $r_p$, and a fiber-coupling lens focuses it to a diameter $d_p$ that should coincide with the fiber’s mode field diameter for optimal fiber coupling. 
	(\textbf{b}) Numerically calculated first-lens collection enhancement from an optimized bullseye grating as a function of the objective’s NA (black). The enhancement is relative to a bare GaAs substrate. Results for a single conventional DBR are also shown (dotted gray). 
	(\textbf{c}) Collection efficiency for the system as a function of $\theta_p$ (black). Objective (dotted gray) and fiber-coupling (dashed gray) components are also shown. The objective NA (0.7) and fiber-coupling optics are held constant. Fiber coupling was optimized around 44$^\circ$, corresponding to NA=0.7. Each component has been individually normalized to unity. 
	}
\label{fig:A2}
\end{figure}

To illustrate the effect, consider the simplified setup shown in Fig. \ref{fig:A2}a. Photons emitted from a point source (``emitter") are collected by an objective (``first lens") with numerical aperture NA and focused to the tip of a single-mode optical fiber using a lens with focal length $f$. Two cases are illustrated: (1) Without a photonic structure (light red region, solid red lines), the maximum angular range over which photons are collected is limited by the objective. At the back aperture, collimated emission has a beam radius $r_0$. A strategic fiber coupling optic focuses this beam tightly to a diameter $d_0$ to match the mode field diameter (MFD) of the single-mode collection fiber. (2) With a photonic structure (darker red region, dotted red lines), emission is concentrated into a smaller angular range with divergence angle $\theta_p$. At the back aperture, the collimated beam thus has a smaller beam radius, $r_p$$<$$r_0$. The same fiber coupling optic focuses this beam to a larger diameter, $d_p$$>$$d_0$, and fiber coupling suffers. 

We first address the impact of the objective. Our experiments (summarized in Fig. \ref{fig:3}) used an objective with an effective NA of approximately 0.25. (Two thick optical windows between the objective and sample perturb the confocal performance of the setup by affecting mode-matching between the collected photons and the single-mode collection fiber. The result is that our objective with nominal NA of 0.7 has an effective NA of approximately 0.25.) In this case, the bullseye grating which concentrates farfield photons well within NA=0.25 (e.g., Fig. \ref{fig:2}c) is obviously beneficial. For objectives with larger NAs, this benefit is expected to be reduced. Fig. \ref{fig:A2}b (black solid curve) quantifies how the first-lens collection enhancement depends on the objective’s NA. Here, the calculated farfield intensity from an optimized bullseye grating is normalized to that from a bulk single-interface GaAs geometry (``Reference system 1"), both for the same NA. For NA$\leq$0.25, first-lens collection enhancements are between 30$\times$ and 100$\times$, similar to those observed in our measurements. Values calculated at NA=1.0 correspond to the ratio of \emph{total} photons emitted into air relative to the substrate. Importantly, the bullseye grating provides approximately 10$\times$ better total photon emission into air. For comparison, calculations for a conventional single-DBR geometry (``Reference system 4") are also shown (gray dotted curve). Bullseye gratings outperform the conventional DBR by at least 2$\times$ over all NAs, and up to 10$\times$ for the smallest NAs.

Fig. \ref{fig:A2}c quantifies how end-to-end efficiency varies with the photonic structure’s emission divergence angle $\theta_p$. The fiber-coupling optic was selected to optimize collection based on the objective’s nominal NA (0.7, corresponding to $\theta$=44$^\circ$; dashed gray curve). As the photonic structure concentrates light into a divergence angle $\theta_p$, the first-lens collection efficiency (dotted gray curve) increases, but the fiber-coupling efficiency decreases. The total collection efficiency (solid black curve) is a product of these two efficiencies and reaches a maximum around $\theta_p$=25$^\circ$ (corresponding to NA=0.24). That is, \emph{total} collection efficiencies can be further improved by choosing new optimal fiber-coupling optics appropriate for the narrower collected beam radius $r_p$.

\bibliography{man}

\end{document}